\documentclass[prd,nofootinbib,superscriptaddress,twocolumn,floatfix]{revtex4}

\usepackage{times} \usepackage{pdfsync}

\newcommand{\ka}{\textsc{k xviii}\xspace}%
\newcommand{\ar}{\textsc{Ar xvii}} %
\newcommand{\ca}{\textsc{Ca xix}} %
\newcommand{\flux}{\,\unit{cts/sec/cm^2}}

\usepackage{amssymb,amsfonts,amsmath,paralist,xspace} %

\usepackage{longtable}%
\usepackage[normalem]{ulem} %
\usepackage{relsize,color,xcolor} %
\usepackage{booktabs} % provides \toprule, \bottomrule, \midrule
\usepackage{units} % Command $\unit[10]{GeV}$ prints correctly 10 GeV. Command
\usepackage{grffile} %
\usepackage{graphicx} %
\graphicspath{{./figures/}} % Search for files also there. Last ``/'' is
\IfFileExists{srcltx.sty}{\usepackage[active]{srcltx}}

\usepackage{etoolbox} %
\preto\tabular{\setcounter{magicrownumbers}{0}}%
\newcounter{magicrownumbers}%
 \newcommand{\be}
{\begin{equation}} \newcommand{\ee} {\end{equation}}   

\def\be{\begin{eqnarray}} \def\ee{\end{eqnarray}} 
\begin{document}

\title{Comment on the paper ``Dark matter searches going bananas: the
  contribution of Potassium (and Chlorine) to the 3.5 keV line'' by T. Jeltema
  and S. Profumo}

\author{A.~Boyarsky$^{1}$, J.~Franse$^{1,2}$, D.~Iakubovskyi$^{3}$, and O.~Ruchayskiy$^{4}$ \\
  $^1${\small Instituut-Lorentz for Theoretical Physics, Universiteit Leiden,
    Niels Bohrweg 2, Leiden, The Netherlands}\\
  $^2${\small Leiden Observatory, Leiden University, Niels Bohrweg 2, Leiden, The Netherlands}\\
  $^3${\small Bogolyubov Institute of Theoretical Physics, Metrologichna
    Str. 14-b, 03680, Kyiv, Ukraine}\\
  $^4${\small Ecole Polytechnique F\'ed\'erale de Lausanne, FSB/ITP/LPPC, BSP, CH-1015, Lausanne, Switzerland}\\
} \date{\today}

\begin{abstract}
  We revisit the X-ray spectrum of the central $14'$ of the Andromeda galaxy,
  discussed in our previous work \cite{Boyarsky:14a}. Recently in
  \cite{Jeltema:14} it was claimed that if one limits the analysis of the data
  to the interval $3-4$ keV, the significance of the detection of the line at
  $3.53$~keV drops below $2\sigma$.  In this note we show that such a
  restriction is not justified, as the continuum is well-modeled as a power
  law up to $8$~keV, and parameters of the background model are well
  constrained over this larger interval of energies. This allows for a
  detection of the line at $3.53$~keV with a statistical significance
  greater than $\sim3\sigma$ and for the identification of several known atomic
  lines in the energy range $3-4$ keV.  Limiting the analysis to the $3-4$~keV
  interval results in increased uncertainty, thus decreasing the significance
  of the detection.  We also argue that, with the M31 data included, a
  consistent interpretation of the $3.53$~keV line as an atomic line of \ka in
  all studied objects is problematic.
\end{abstract}

\maketitle
Earlier this year,
two independent groups~\cite{Bulbul:14,Boyarsky:14a} reported a detection
of an unidentified X-ray line at an energy of $\sim 3.53$~keV in the long-exposure
X-ray observations of a number of dark matter-dominated objects. 
The possibility that this spectral feature may be the 
signal from decaying dark matter has sparked a lot of interest in the
community as the signal has passed a number of ``sanity checks''
expected for a dark matter decay signal: it scales correctly between galaxy
clusters, the Andromeda galaxy, the Milky Way center and the upper bound from
non-detection in the blank sky data~\cite{Boyarsky:14a,Boyarsky:14b},
and also between
different subsamples of clusters~\cite{Bulbul:14}.  The signal has radial
surface brightness profiles in the Perseus cluster, Andromeda
galaxy~\cite{Boyarsky:14a} and in the Milky Way~\cite{Boyarsky:14b} that are
consistent with our expectations about the dark matter distribution in these
objects.

Recently, the authors of \cite{Jeltema:14} have argued that if one restricts
the modeling of the emission of the central part of M31 to the energy range
$3-4$~keV, and uses a single \texttt{powerlaw} as a model of the continuum,
the significance of the detection of the line at $3.53$~keV in the spectrum of
M31 drops below $2\sigma$. They also argued that when one ignores the detection in
M31, the line in the spectra of the galaxy clusters and of the Galactic Center
can be explained by an atomic transition in the \ka ion, provided one also
assumes both an abundance of \ka and a set of physical conditions in these
objects that are hard to exclude.

\emph{In this note we show that restricting the analysis of the M31 spectrum
  to $3-4$~keV is not justified. The continuum is well modelled by a
  power law model up to 8 keV and the parameters of this model are well
  constrained at this wider interval. Limiting the analysis to $3-4$~keV only
  results in increased uncertainty and, although the flux in the
  $3.53$~keV line is consistent with the one reported in~\cite{Boyarsky:14a},
  the significance of its detection is naturally smaller on the $3-4$~keV than
  on the whole $2-8$~keV interval, where the astrophysical background is
  better constrained.  We also argue that with the M31 data included, the
  interpretation of the $3.53$~keV line as a \ka line in several studied objects
  together is problematic.}

We start by repeating the analysis of \cite{Jeltema:14}: we fit the M31
spectrum over the interval 3--4~keV with a single \texttt{powerlaw} (in
order to avoid having to model the instrumental background, we subtract
it from our spectra.)\footnote{The spectral
  modeling has been performed with the X-Ray Spectral Fitting Package
  \texttt{Xspec}~\cite{Arnaud:96} v.12.8.0.}  The fit is good ($\chi^2
= 22.4$ for 27 d.o.f.).\footnote{Unlike \cite{Jeltema:14} we have binned the
  spectrum by 60~eV (as in \cite{Boyarsky:14a}) to make bins roughly
  statistically independent. We verified that our conclusion does not change
  for finer binning.}  The parameters of the \texttt{powerlaw} are: PL index
$1.65 \pm 0.05$ (3\% relative error), and PL norm $(1.19\pm 0.07)\times 10^{-3}\flux/\mathrm{keV}$ at 3.5
keV (the relative error being 6.3\%).  An additional line is detected against this
continuum at energy $3.53$~keV and with normalization $(2.7\pm 1.5)\times
10^{-6}\flux$ (less than $2\sigma$ significance, $\Delta \chi^2 = 3.4$ when
adding this line). Thus, we have reproduced both the flux and the
significance reported in \cite{Jeltema:14}.

\begin{figure*}[!t]
  \includegraphics[width=0.49\textwidth,angle=0]{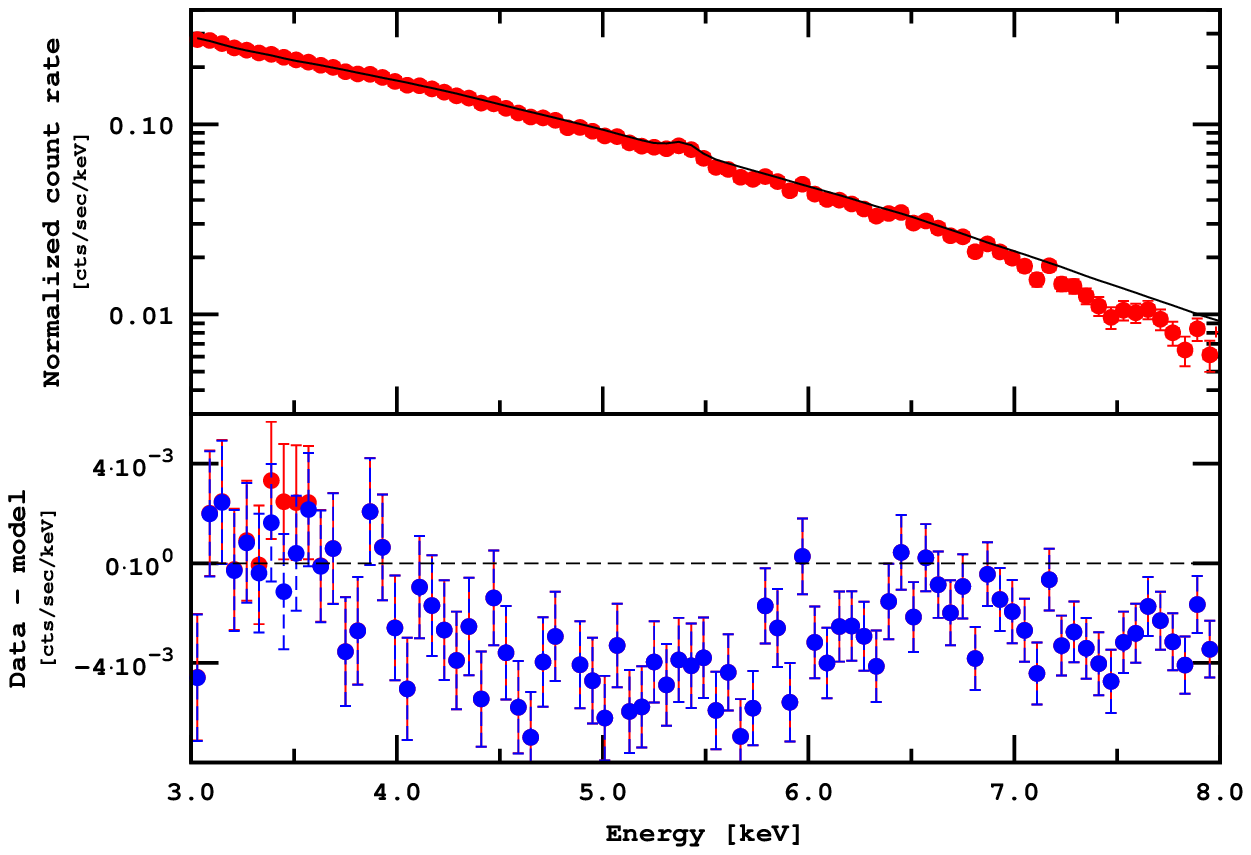}
  \includegraphics[width=0.49\textwidth]{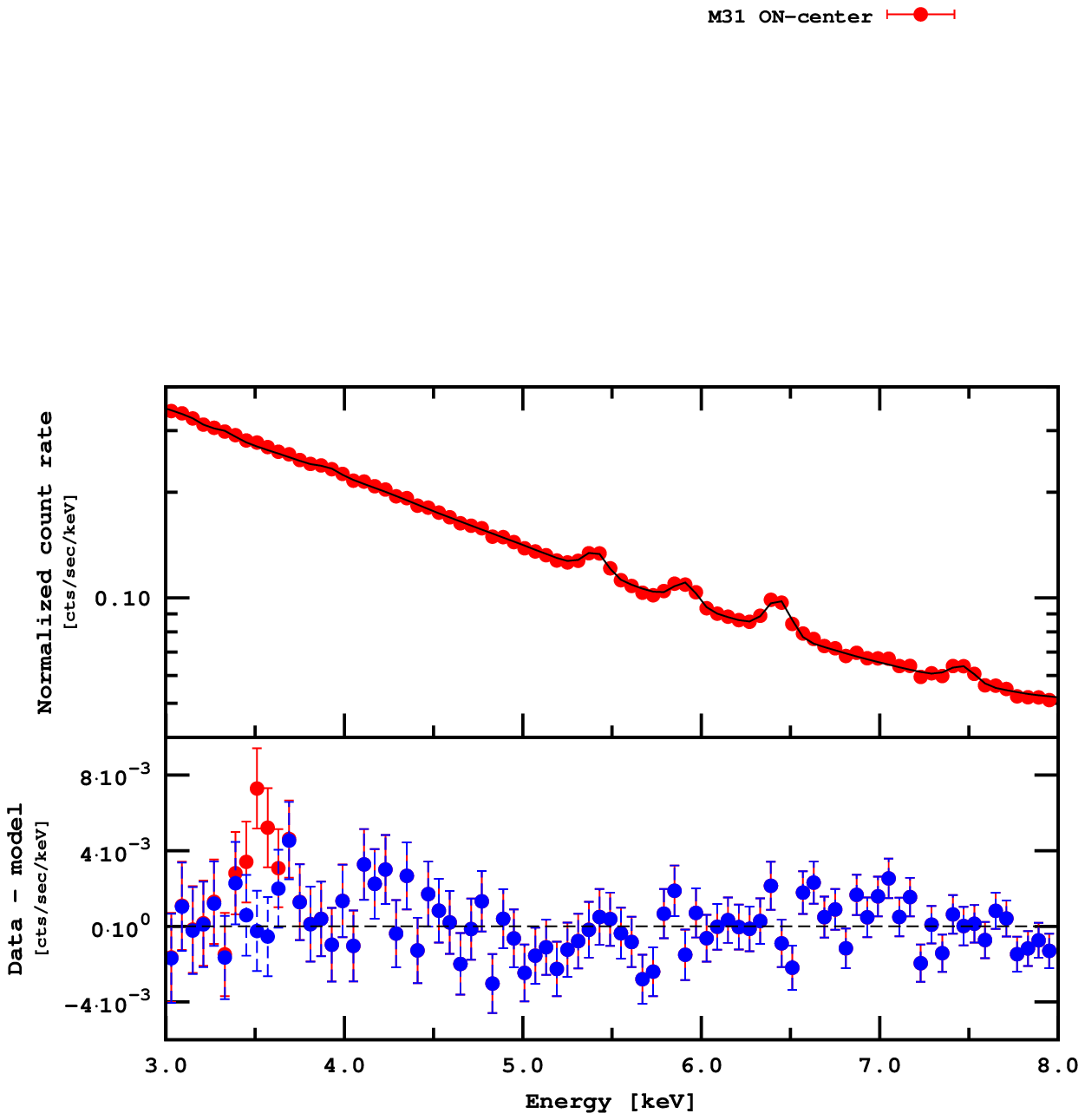}
  \caption{\textit{Left panel:} Best fit powerlaw, as determined over the
    3--4~keV interval in \protect\cite{Jeltema:14}, extended to higher
    energies. Such a powerlaw significantly overpredicts the count rate in all
    bins at $E>4$~keV. A wide group of near-zero residuals around $6.5$~keV
    corresponds to the complex of lines of iron and other elements
    (unmodeled). \textit{Right panel:} The data over a wider range of energies
    processed and fitted in \protect\cite{Boyarsky:14a}) with the line at
    $3.53$ keV unmodeled. Other lines in the range $3-4$~keV are included in
    the model.}
  \label{fig:PL_overpredict}
\end{figure*}

However, once we extend the powerlaw obtained over the interval $3-4$~keV to
higher energies, we see that it \emph{significantly overpredicts} the count
rate in all energy bins above $4$~keV as Fig.~\ref{fig:PL_overpredict}
demonstrates.

Let us now compare this result with the fit over the whole interval 2--8~keV
(as in Ref.~\cite{Boyarsky:14a}). The wider range of energies allows us to
determine the parameters of the \texttt{powerlaw} with better precision: PL
index $1.71\pm 0.01$ (0.5\% relative error), PL norm = $(1.18\pm 0.01) \times
10^{-3}\flux/\mathrm{keV}$ at 3.5~keV (0.7\% relative error).  The improvement
of the quality of fit when adding the line around $3.53$~keV was $\Delta \chi^2
= 13$ (which is about $3\sigma$ for 2 degrees of freedom: position and
normalization of the line). This is the most significant feature in the
3--4~keV range.  In Table~\ref{tab:m31-line-properties} we list all the lines
detected in the interval 3--4~keV with significance more than
$1\sigma$. Unlike the results of \cite{Jeltema:14} (see Fig.~3 therein), the
lines at $3.91$~keV (complex of \ca\ and \ar\ lines) have also been
detected in this case with the significance above $2\sigma$. In the case of
the fit over the $3-4$~keV range, these lines were partially compensated by
the \texttt{powerlaw} continuum.  In addition, the parameters of the continuum
as determined over the narrow range of energies naturally suffer from larger
errors (around $3-6\%$ for the fit over $3-4$~keV interval vs.\ $0.5-0.7\%$ for
the fit of \cite{Boyarsky:14a}).  As the flux in the line in question is about
$4\%$ of the continuum at these energies, the parameters of the background
model should be determined with a precision greater than that in order to
reliably detect such a weak line. This explains the reduced best-fit flux and
the diminished significance of the line at $3.53$~keV.

\begin{table}[!b]
  \centering
  \caption{Position and flux of lines found in the central part of M31~\protect\cite{Boyarsky:14a}, together with 
    $1\sigma$ error ranges.}\label{tab:m31-line-properties}
 \vspace*{1ex}
 \begin{tabular}{lcc}
   \hline
   Line                   & Position, keV             & Flux, ph/sec/cm$^2$ \\
   \hline
   {\ar}/\textsc{S~XV}           & $3.14\pm 0.04$ & $2.3\pm 1.4 \times 10^{-6}$ \\
   {\textsc{Ar XVIII}/\textsc{S XVI}/\textsc{Cl XVI}}  & $3.37\pm 0.03$ & $3.6\pm 1.4 \times 10^{-6}$ \\
   {\ar/\ca}         & $3.91\pm 0.02$ & $4.3\pm 1.3 \times 10^{-6}$ \\
   DM line candidate        & $3.53\pm 0.03$ & $4.9^{+1.6}_{-1.3} \times 10^{-6}$ \\
   \hline 
 \end{tabular}
\end{table}

\begin{figure}[!t]
  \centering
  \includegraphics[width=0.5\textwidth]{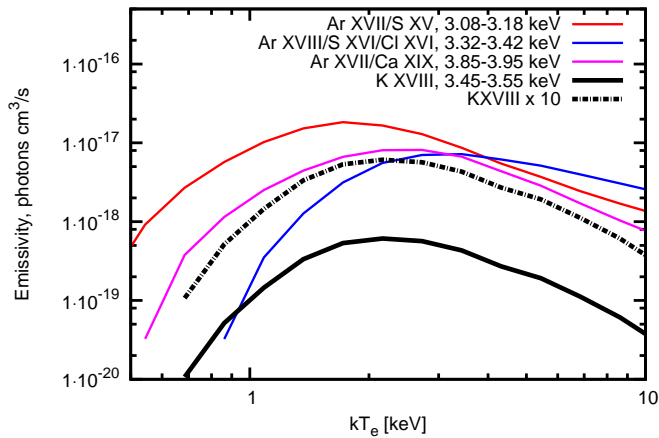}
  \caption{Emissivities of different line complexes as a function of plasma
    temperature (assuming solar abundances of all elements). The emissivity of
    the \ka line complex near 3.5~keV is at least an order of magnitude below those of the other
    complexes. The data is from AtomDB v2.0.2~\protect\cite{AtomDB}.}
  \label{fig:emissivities}
\end{figure}

Finally, we make the observation that complexes of argon, calcium and sulphur
at energies 3.14~keV, 3.37~keV and 3.91~keV (of which only the 3.91~keV
complex is detected at more than $2\sigma$) have fluxes \emph{lower} than that
of the unidentified spectral feature at $3.53$~keV. This \emph{challenges} the
interpretation of the feature as a \ka complex. Indeed, according to AtomDB
v2.0.2~\cite{AtomDB} \ka emissivity is at least an order of magnitude lower
than emissivities of the complexes in the intervals $3.85-3.95$~keV and
$3.08-3.18$~keV (see Fig.~\ref{fig:emissivities} based on the data
from~\cite{AtomDB}). This relation between emissivities is based on the
assumption of solar abundances for these elements. To change this conclusion a
strongly super-solar abundance of \ka would be required.

\emph{In conclusion:} the line $3.53$~keV is detected at $\sim 3 \sigma$ level
in the spectrum of the Andromeda galaxy against a background model with the
continuum component constrained at the 2--8~keV interval as in
\cite{Boyarsky:14a}. Fitting the
data in the much narrower
$3-4$~keV range reduces the significance of the detection as the sensitivity
likewise reduces with less data, however this does not contradict the flux in the line
detected in \cite{Boyarsky:14a}.  The fit over the narrow interval of
energies, as performed in \cite{Jeltema:14}, provides a best fit value of the
slope of the power law background that systematically over-predicts the value
of the flux above 4 keV, and is therefore significantly ruled out by the whole
spectrum.

The observation of the line at $3.53$~keV in the center of M31 is in stark
contradiction with its interpretation as a \ka atomic transition -- it would
require an extremely super-solar abundance of \ka \emph{and} a super-solar ratio of
abundance of \ka relative to \ar\ and \ca.  The presence of this line in
different types of objects -- galaxy clusters, M31, and the Galactic
Center -- makes it
challenging to explain all these signals together by emission from \ka, even if
this interpretation is hard to exclude from the GC data only.

\textbf{Acknowledgments.} We acknowledge useful discussions with T.~Jeltema
and thank Mark Lovell for providing comments on the manuscript.  The work of
D.~I. was supported in part by the Swiss National Science Foundation grant
SCOPE IZ7370-152581, the Program of Cosmic Research of the National Academy of
Sciences of Ukraine, the State Programme of Implementation of Grid Technology
in Ukraine and the grant of President of Ukraine for young scientists.  The
work of J.~F. was supported by the De Sitter program at Leiden University with
funds from NWO. This research is part of the ``Fundamentals of Science''
program at Leiden University.

\bibliography{preamble,astro,combined_numsm,combination_bib,dmline-refs}

\end{document}